\documentclass[aps,prd,showpacs,amsmath,amssymb,twocolumn,superscriptaddress,preprintnumbers,nofootinbib]{revtex4-1}
\pdfoutput=1
\usepackage{graphicx}
\usepackage{amsthm}
\usepackage{amsmath}
\usepackage{graphicx,subfigure}
\usepackage{dcolumn}
\usepackage{bm}
\usepackage{slashed}
\usepackage{calligra}
\DeclareMathAlphabet{\mathcalligra}{T1}{calligra}{m}{n}
\DeclareFontShape{T1}{calligra}{m}{n}{<->s*[2.5]callig15}{}

\newcommand{\be}{\begin{eqnarray}}
\newcommand{\ee}{\end{eqnarray}}
\newcommand{\bea}{\begin{eqnarray}}
\newcommand{\eea}{\end{eqnarray}}

\newcommand{\MeV}{{~\rm MeV}}
\newcommand{\GeV}{{~\rm GeV}}

\newcommand{\dPS}{{\rm dPS}}
\newcommand{\mupmum}{\mu^+\mu^-}

\newcommand{\zprime}{{\rm Z^\prime}}
\newcommand{\mzp}{m_{\zprime}}

\newcommand{\gp}{g'}

\newcommand{\CV}{C_{\rm _V}}
\newcommand{\CA}{C_{\rm _A}}
\newcommand{\thetaW}{\theta_{\rm _W} }
\newcommand{\GF}{G_{\rm F}}

\begin{document}
\title{Neutrino Trident Production: A Powerful Probe of New 
Physics with Neutrino Beams}
\author{Wolfgang Altmannshofer}
\affiliation{Perimeter Institute for Theoretical Physics 31 Caroline St. N, Waterloo, Ontario, Canada N2L 2Y5.}
\author{Stefania Gori}
\affiliation{Perimeter Institute for Theoretical Physics 31 Caroline St. N, Waterloo, Ontario, Canada N2L 2Y5.}
\author{Maxim Pospelov}
\affiliation{Perimeter Institute for Theoretical Physics 31 Caroline St. N, Waterloo, Ontario, Canada N2L 2Y5.}
\affiliation{Department of Physics \& Astronomy, University of Victoria, Victoria, BC, V8P 5C2, Canada}
\author{Itay Yavin}
\affiliation{Perimeter Institute for Theoretical Physics 31 Caroline St. N, Waterloo, Ontario, Canada N2L 2Y5.}
\affiliation{Department of Physics \& Astronomy, McMaster University 1280 Main St. W. Hamilton, Ontario, Canada, L8S 4L8.}

\begin{abstract}

The production of a $\mupmum$ pair from the scattering of a muon-neutrino off the Coulomb field of a nucleus, known as neutrino trident production, is a sub-weak process that has been observed in only a couple of experiments. As such, we show that it constitutes an exquisitely sensitive probe in the search for new neutral currents among leptons, putting the strongest constraints on  well-motivated and well-hidden extensions of the Standard Model gauge group, including the one
coupled to the difference of the lepton number between the muon and tau flavor, $L_\mu-L_\tau$. The 
new gauge boson, $\zprime$,
 increases the rate of neutrino trident production by inducing additional $(\bar \mu \gamma_\alpha \mu) (\bar\nu\gamma^\alpha \nu)$
 interactions, which interfere constructively with the Standard Model contribution.
Existing experimental  results put significant restrictions on 
the parameter space of any model coupled to muon number $L_\mu$, 
and disfavor a putative resolution to the muon $g-2$ discrepancy via the loop of $\zprime$ for any mass 
 $m_\zprime \gtrsim 400$~MeV. The reach to the models' parameter space can be widened 
with future searches of the trident production at high-intensity neutrino facilities such as the LBNE.

\end{abstract}

\pacs{12.60.Cn, 13.15+g, 25.30.Pt}

\maketitle

{\em Introduction.} The Standard Model (SM) gauge group is one of its most important defining features, giving fully adequate description to all electroweak and strong interaction 
phenomena. However, there is no reason to believe that the $SU(3)\times SU(2)\times U(1)$ structure is final, and its extensions, both at some high-energy scale, and 
 at low energy, have been discussed in the literature, and subjected to a multitude of experimental searches. New Abelian gauge groups, $U(1)_X$, 
are of particular interest, as many top-down approaches predict their possible existence \cite{Langacker:2008yv}. The simplest possibility is a new gauge group 
coupled to the SM via a gauge invariant renormalizable portal, known as kinetic mixing~\cite{Holdom:1985ag}, 
while the SM fields maintain their complete neutrality with respect to $U(1)_X$. There are also well-known possibilities in which the SM fields carry a charge under a 
new force. 
The requirement that such theories are valid up to very high-energy scales singles out the anomaly-free combinations of gauged $X=y B -\sum_i x_i L_i$ number. Here $B$ is the baryon number, $L_i$ are individual lepton flavor numbers, and $y, x_i$ are the constants related by the anomaly-free requirement, $3y= x_e+x_\mu+x_\tau$.
Models with $y,x_e \neq 0$ are generally well-constrained by 
electron and proton colliders,  as well as neutrino scattering experiments. Yet, there is one combination,
 $y=x_e=0;~x_\mu=-x_\tau$, resulting in a new force associated with muon number minus tau number ($L_\mu-L_\tau$)~\cite{He:1990pn,He:1991qd}
that  is difficult to probe since it would affect only neutrinos and the unstable leptons. 

One robust consequence of a new force that couples to muons via a vector portal, either $L_\mu$ and/or kinetic mixing with the photon, 
is the additional positive contribution to the muon anomalous magnetic moment $a_\mu$. Since there exists a long-standing discrepancy
in the muon $g-2$ between experiment and SM prediction at the $\sim3.5\sigma$ level, a possible increase of $a_\mu$ by $\sim3\times 10^{-9}$, 
due to a new vector force, may solve this problem. Until recently the existing constraints were 
sufficiently weak to afford the possibility of a ``dark force'' resolution to the $g-2$ discrepancy over much of the model parameter space~\cite{Baek:2001kca,Ma:2001md,Gninenko:2001hx,Pospelov:2008zw,Heeck:2011wj,Harigaya:2013twa}.  By now, the kinetically mixed $\zprime$ (also known as ``dark photon'') has been subjected to a multitude of experimental tests that almost entirely rule out the region of parameter space relevant to muon $g-2$~\cite{Essig:2013lka,Merkel:2014avp}. Could comparatively strong bounds be found for models with the gauged $L_\mu$? 

In this \textit{Letter}, we show that any model based on gauged muon number, $L_\mu$, is significantly 
restricted by the rare SM process of {\it{neutrino trident production}}: 
the production of a $\mupmum$ pair from the scattering of a muon-neutrino with heavy nuclei. 
The observation of this process in neutrino beam experiments at levels consistent with the SM strongly constrains 
contributions from a new force~\cite{Altmannshofer:2014cfa}. 
Perhaps more importantly, we show that future neutrino beam facilities, such as LBNE, may be able to search for
such forces in yet unconstrained regions of the parameter space. Our present work extends and generalizes the arguments given in~\cite{Altmannshofer:2014cfa} for a heavy $\zprime$, and in particular rules out such force as a solution of the muon $g-2$ discrepancy over 
a large portion of the relevant parameter space, $\mzp \gtrsim 400\MeV$. Also, given the importance of this process for new physics, we recalculate the 
rate of the neutrino trident production in the SM.

{\em Muonic tridents in the SM and beyond.} To be specific, and to take the least constrained case, we concentrate on a $\zprime$ boson coupled to $L_\mu-L_\tau$,  
\be
\label{Zprime}
\mathcal{L}_{\zprime} &=& -\tfrac{1}{4}\left(\zprime \right)_{\alpha\beta} \left(\zprime \right)^{\alpha\beta} + \tfrac{1}{2} \mzp^2 {\rm Z'}_{\alpha}{\rm Z'}^{\alpha}
\\ 
\nonumber
&& + \gp \zprime_\alpha \Big( \bar{\ell}_2 \gamma^\alpha \ell_2 -\bar{\ell}_3 \gamma^\alpha \ell_3  + \bar{\mu}_{_R} \gamma^\alpha \mu_{_R} - \bar{\tau}_{_R} \gamma^\alpha \tau_{_R} \Big)~.
\ee
Here, $\gp$ is the $U(1)_X$ gauge coupling, the field-strength is $\left(\zprime \right)_{\alpha\beta} = \partial_\alpha \zprime_\beta - \partial_\beta \zprime_\alpha$, the electroweak doublets associated with left-handed muons and taus are $\ell_2 = (\nu_\mu , \mu_{_L})$ and $\ell_3 = (\nu_\tau , \tau_{_L})$, and the right-handed electroweak singlets are $\mu_{_R}$ and $\tau_{_R}$. The origin of the vector boson mass is not directly relevant for our work, and thus we suppress any additional pieces in (\ref{Zprime}) related to the corresponding Higgs sector. 

\begin{figure}[t]
\centering
\includegraphics[width=0.34\textwidth]{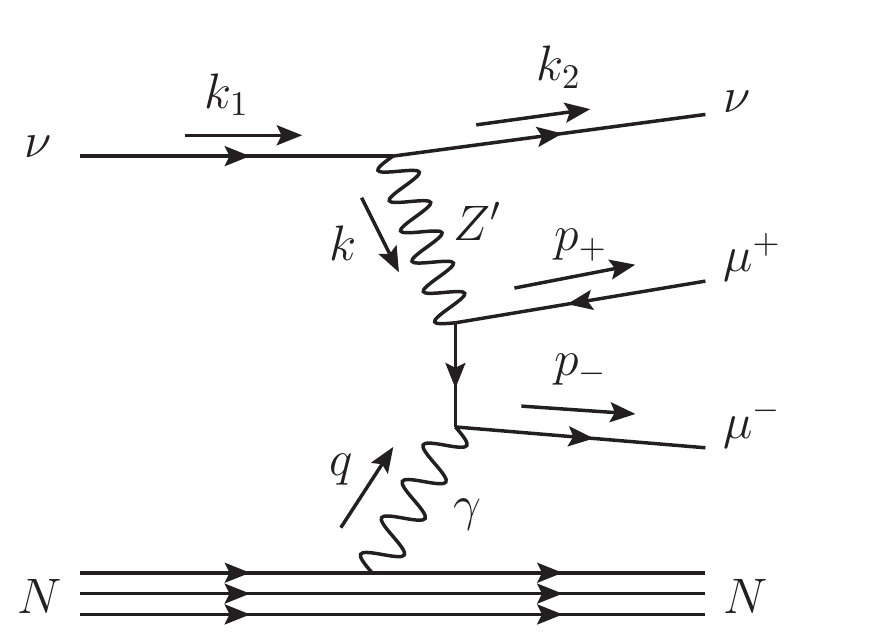}%
\caption{The leading order contribution of the $\zprime$ to neutrino trident production (another diagram with $\mu^+$ and $\mu^-$ reversed is not shown). Other contributions at the same order in $\gp$ are further suppressed by the Fermi scale.
}
\label{fig:trident_diagram}
\end{figure}

This model contributes to the neutrino trident production at lowest order through the diagram shown in Fig.~\ref{fig:trident_diagram}. This contribution interferes with the SM contribution coming from $W^\pm/Z$ exchange. In order to gain insight into the different contributions, in what follows we provide analytical results using the equivalent photon approximation (EPA)~\cite{vonWeizsacker:1934sx,Williams:1934ad}. Under the EPA, the full cross-section of a muon-neutrino scattering with a nucleus N is related to the cross-section of the neutrino scattering with a real photon through,
\be
\label{eqn:EPA}
\hspace{-4mm} \sigma(\nu_\mu  N \to \nu_\mu N \mupmum) = \int \sigma(\nu_\mu  \gamma \to \nu_\mu \mupmum)~P(s,q^2) ~.
\ee 
Here, $P(q^2,s)$ is the probability of creating a virtual photon in the field of the nucleus N with virtuality $q^2$ which results in the energy being $\sqrt{s}$ in the center-of-mass frame of the incoming neutrino and a real photon. This probability is given by~\cite{Belusevic:1987cw}
\be
\label{eqn:probability_for_virtual_photon}
P(q^2,s)= \frac{Z^2 e^2}{4\pi^2}  \frac{ds}{s} \frac{dq^2}{q^2} F^2(q^2)~,
\ee 
where $Ze$ and $F(q^2)$ are the charge and the electromagnetic form-factor of the nucleus, respectively. The integral over $s$ is done from $4m^2$ to $2E_\nu q$, with the muon mass $m$ and the neutrino energy $E_\nu$. The $q$ integral has a lower limit of $4m^2/(2E_\nu)$ and the upper limit is regulated by the exponential form-factor. We thus concentrate on the computation of the cross-section $ \sigma(\nu_\mu  \gamma \to \nu_\mu \mupmum)$. 
Computations of the full $\nu_\mu  N \to \nu_\mu N \mupmum$ process have been performed in~\cite{Czyz:1964zz,Fujikawa:1971nx,Lovseth:1971vv,Koike:1971tu,Koike:1971vg,Brown:1973ih} in the context of the V-A theory and of the SM.

We begin with the differential cross-section for the $\nu\gamma\to \nu\mu^+\mu^-$ sub-process associated with a pure V-A charged interaction between neutrinos and muons.
It is given symbolically by
\bea
\label{master}
d\sigma =  \frac{1}{2s} \dPS_3 \left(\frac{1}{2} \sum_{\rm pol} |M_1 M_2|^2 \right) \frac{\GF^2 e^2}{2} ~,
\eea
where $G_F = \sqrt{2}g^2/(8M_{_W}^2)$ is the Fermi constant. The 3-body phase-space (with correction of a typo in the corresponding expression of ref.~\cite{Vysotsky:2002ix})
is given by
\be
\dPS_3 = \frac{1}{2}\frac{1}{(4\pi)^2}\frac{dt}{2s} \frac{d\ell}{2\pi} ~v \frac{d\Omega'}{4\pi} ~,
\ee
where $\ell = (p_++p_-)^2$ is the square of the invariant mass of the $\mupmum$ pair, $\Omega'$ is the solid angle with respect to the photon four-vector in the $\mupmum$ rest-frame, $v=\sqrt{1-4m^2/\ell}$ is the velocity of each muon in that frame, and $t \equiv 2k\cdot q$. $M_1$ and $M_2$
in (\ref{master}) are the neutrino and the muon-pair blocks in the amplitude, that form the total amplitude according to $M = \frac{\GF e}{\sqrt{2}}  M_1 M_2$.
The factor of $1/2$ in (\ref{master}) originates from the average over the incoming photon polarizations.

Using  $M_{1,2}$ explicitly, and summing over spins and polarizations, we get (in agreement with result of ref. \cite{Belusevic:1987cw})
\bea
\label{lengthyformula}
&~&\frac{1}{2} \sum_{\rm pol} |M_1 M_2|^2 \equiv 512~ \left| \mathcal{M}_{\rm V-A} \right|^2 \simeq 512 \times \Bigg( \\ \nonumber
&~&\frac{ (k_1\cdot p_+) (q\cdot k_2) (q\cdot p_-)}{A^2} +\frac{ (k_2\cdot p_-) (q\cdot k_1) (q\cdot p_+)}{B^2}  \\ \nonumber
&+&\frac{2 (k_1\cdot p_+)( k_2\cdot p_-)( p_+\cdot p_-)}{A B} -\frac{ (k_2\cdot p_-) (p_+\cdot p_-) (q\cdot k_1)}{A B} \\ \nonumber
&-&\frac{ (k_1\cdot p_+) (p_+\cdot p_-) (q\cdot k_2)}{A B} -\frac{ (k_1\cdot p_+)( k_2\cdot p_-) (q\cdot p_-)}{A B} \\ \nonumber
&+&\frac{ (k_1\cdot p_+) (k_2\cdot p_+)( q\cdot p_-)}{A B} +\frac{ (k_1\cdot p_-) (k_2\cdot p_-) (q\cdot p_+)}{A B} \\ \nonumber
&-&\frac{ (k_1\cdot p_+)( k_2\cdot p_-) (q\cdot p_+)}{A B} \Bigg) ~,
\eea
 where $A = (p_- - q)^2 - m^2$ and $B = (q-p_+)^2 - m^2$.
The result for the full SM contribution together with the $\zprime$ vector-boson exchange can be obtained from the V-A matrix-element contribution, if we neglect terms proportional to the muon mass. The full square of the matrix-element is defined as in Eq.~(\ref{lengthyformula}) but with,
\bea
\label{eqn:fullME}
\frac{1}{2} \sum_{\rm pol} |M_1 M_2|^2 &=&  512 ~ \left| \mathcal{M}_{\rm V-A} \right|^2 \times \frac{1}{2} \Bigg( \CV^2 + \CA^2 
\\ \nonumber
&& \hspace{-16mm} \left. - 2 \CV \CV^{\rm (Z')} \frac{\mzp^2}{k^2-\mzp^2}  +  \left(\CV^{\rm (Z')} \frac{\mzp^2}{k^2-\mzp^2} \right)^2  \right)~.
\eea
Here, $k$ is the momentum of the exchanged $\zprime$ and the SM coefficients of the vector and axial-vector currents in the interaction of muon-neutrinos with muons are 
$\CV = \frac{1}{2} + 2 \sin^2\thetaW$, $\CA = \frac{1}{2}$,
with $\thetaW$ being the weak mixing angle. 
The second line in Eq.~(\ref{eqn:fullME}) features the $\zprime$ contribution with the vector-current coefficient defined as,
\be
\CV^{(\rm Z')} = 4 \frac{M_W^2}{\mzp^2}\frac{\gp^2}{g^2} = \frac{v_{\rm{SM}}^2}{v_{_{\zprime}}^2} ~,
\ee
where $v_{\rm{SM}} = 246$~GeV is the SM Higgs vacuum expectation value and $v_{_{\zprime}} = \mzp/\gp$.
 
Next we consider the phase-space integration. The total cross-section is obtained by 
integrating over the entire solid angle $\Omega'$,  $\ell < t < s$, and $4m^2 < \ell < s$. 
The integration over phase-space is best done first over the solid angle, 
then over $t$ and $\ell$ (see also ref.~\cite{Vysotsky:2002ix}). Keeping only leading log 
terms in the muon mass we find the following expression for the inclusive 
SM cross-section,
\be
\hspace{-4mm}
\sigma^{\rm (SM)} \simeq \frac{1}{2}\left( \CV^2 + \CA^2\right) \frac{2\GF^2 \alpha ~s}{9\pi^2}\left(\log\left(\frac{s}{m^2}\right) - \frac{19}{6} \right) ~.
\ee
The destructive interference between the charged and neutral vector-boson contributions leads 
to a reduction of about 40\% of the SM cross-section compared to the pure V-A theory. Our results corrects a missing factor of 2 
in the corresponding expression in ref.~\cite{Belusevic:1987cw}.

In general we can write
\be
\sigma^{\rm (SM+\zprime)} = \sigma^{\rm (SM)} + \sigma^{\rm (inter)} + \sigma^{\rm (\zprime)}~,
\ee
where the second term is the interference between the SM and the $\zprime$ contributions. In the heavy mass limit, $\mzp \gg \sqrt{s}$ this can be expressed concisely as~\cite{Altmannshofer:2014cfa}
\be
\label{eqn:sigma_tot_high_mass}
\frac{\sigma^{\rm (SM+\zprime)} }{\sigma^{\rm (SM)} } \simeq \frac{1 + \left( 1 + 4 \sin^2\thetaW + 2v_{\rm{SM}}^2/v_{_{\zprime}}^2 \right)^2 }{1 + \left( 1 + 4 \sin^2\thetaW \right)^2} ~.
\ee
This expression also holds for the differential cross-section in this limit, up to muon mass corrections. 

In the limit of light $\zprime$, $\mzp \ll \sqrt{s}$ the expression is more complex. In the leading log approximation, the interference term is given by
\bea
\sigma^{\rm (inter)} &\simeq& \frac{\GF}{\sqrt{2}} \frac{\gp^2  \CV \alpha}{3 \pi^2}~ \log^2\left(\frac{s}{m^2}\right)~.
\eea
The $\zprime$ contribution alone, for $m \ll m_\zprime \ll \sqrt{s}$, is
\bea
\sigma^{\rm (Z')} &\simeq& \frac{1}{m_\zprime^2} ~\frac{\gp^4 \alpha}{6  \pi ^2} ~ \log\left(\frac{m^2_\zprime}{m^2}\right)
~,
\eea
while for $m_\zprime \ll m \ll \sqrt{s}$ it is
\bea
\sigma^{\rm (Z')} &\simeq& \frac{1}{m^2} ~\frac{7\gp^4 \alpha}{72  \pi ^2}  ~ \log\left(\frac{m^2}{m_\zprime^2}\right)
~.
\eea
As can be expected, at high $\mzp$ the $\zprime$ contribution is additive with respect to the SM one (as shown in Eq.~(\ref{eqn:sigma_tot_high_mass})) and 
decouples as $\mzp^{-2}$. For light $\zprime$, on the other hand, the cross-section is only log sensitive to $\mzp$ and the center of mass energy of the event.

To get the total $\nu_\mu  N \to \nu_\mu N \mupmum$ cross-section, the real-photon contribution can be easily integrated against the Weizs{\"a}cker-Williams probability distribution function, Eq.~(\ref{eqn:EPA}), in
$4m^2 < s < 2E_\nu q$ and $4m^2/(2E_\nu) < q < \infty$, with the $q$ integral regulated by the form factor . 
Using a simple exponential form factor, we find good agreement between our results from the EPA and a direct numerical calculation of the full process following~\cite{Lovseth:1971vv}. As a cross check we also reproduced the trident cross sections reported in~\cite{Lovseth:1971vv,Brown:1973ih}, for V-A theory and for the SM, for various neutrino energies, using both the EPA and the numerical calculation. For large $\mzp$ the relative size of the $\zprime$ contribution is independent of the neutrino energy. For low $\mzp$ on the other hand, lower neutrino energies lead to an enhanced sensitivity to the $\zprime$. Since the experimental searches employed a variety of kinematical cuts, in determining the sensitivity to the $\{g',m_\zprime\}$ parameter space we use full numerical results for the phase-space integration rather than analytic approximations and keep the full dependence on the muon mass.

\begin{figure}[t]
\centering
\includegraphics[width=0.4\textwidth]{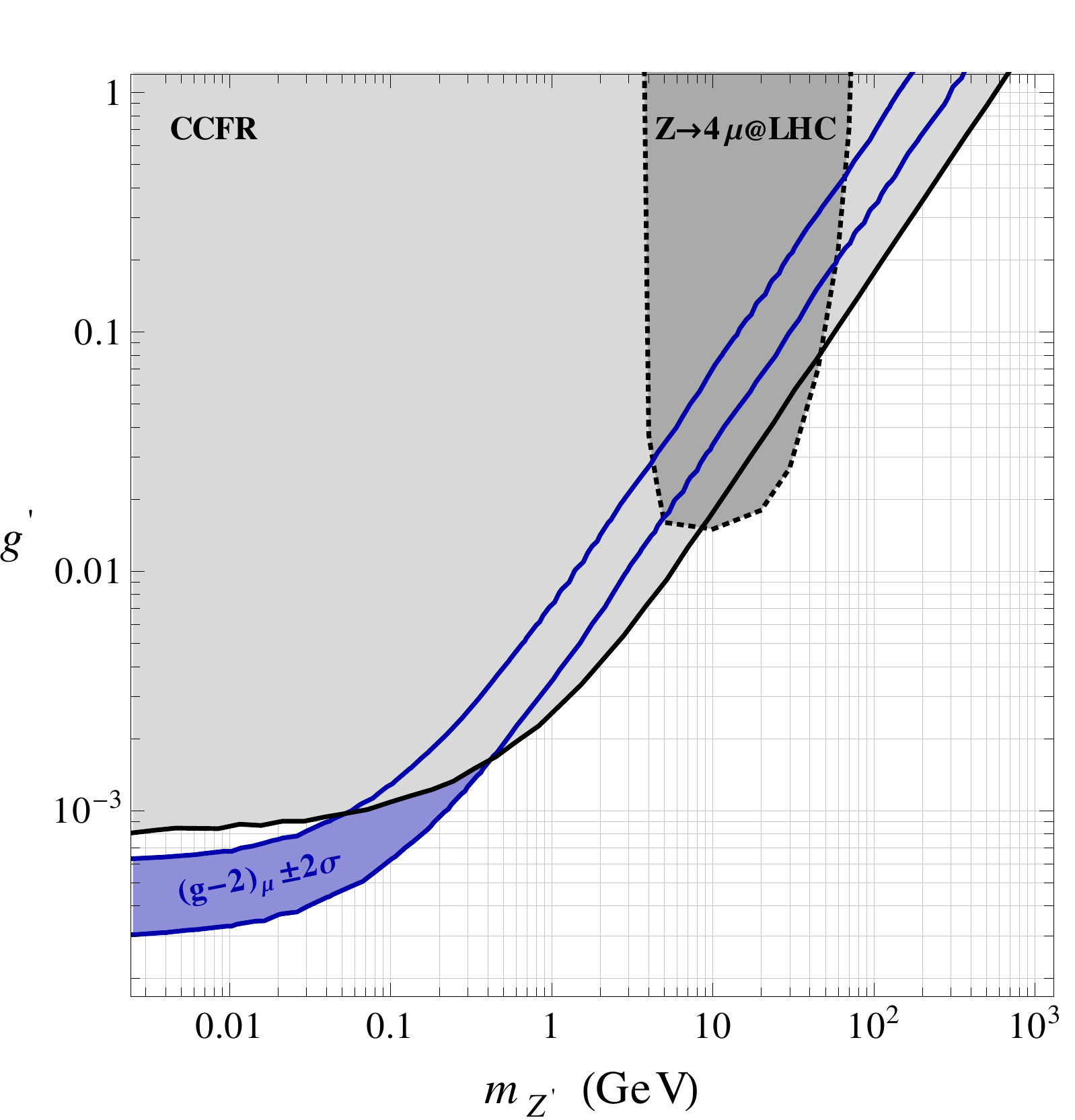} \\
\caption{Parameter space for the $\zprime$ gauge boson. The light-grey area is excluded at 95\%~C.L. by the CCFR measurement of the neutrino trident cross-section. The grey region with the dotted contour is excluded by measurements of the SM $Z$ boson decay to four leptons at the LHC~\cite{CMS:2012bw,Aad:2014wra}. The purple (dark-grey) region is favored by the discrepancy in the muon g-2 and corresponds to an additional contribution of $\Delta a_\mu = (2.9 \pm 1.8) \times 10^{-9}$ to the theoretical value~\cite{Jegerlehner:2009ry}. 
}
\label{fig:constraints}
\end{figure}

\begin{figure}[t]
\centering
\includegraphics[width=0.4\textwidth]{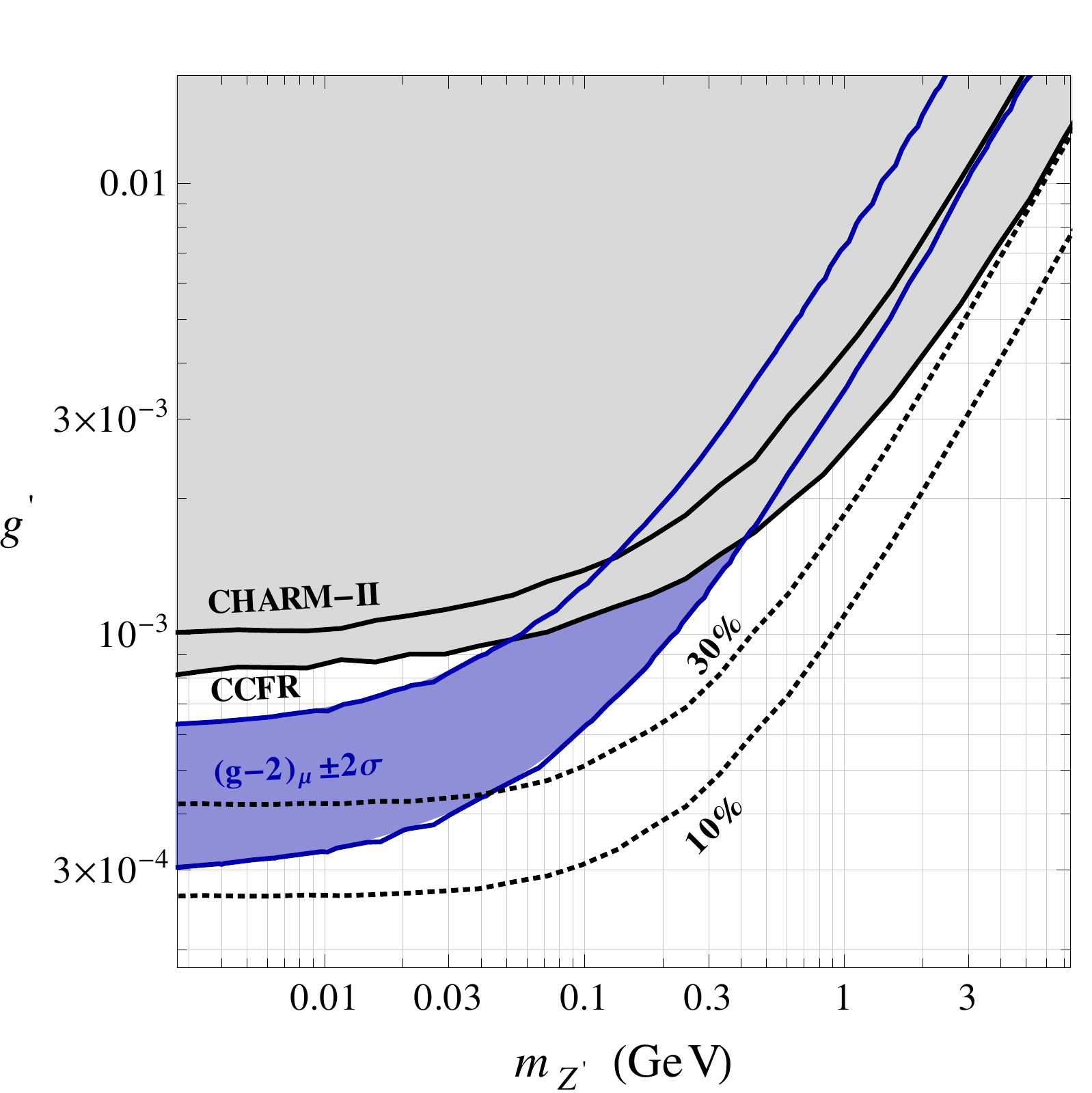}%
\caption{Same as Fig.~\ref{fig:constraints} but focusing on the low mass region. 
Constraints from CHARM-II and CCFR, Eqs.~(\ref{eq:CHARM}) and~(\ref{eq:CCFR}) are shown separately. We do not attempt a statistical combination of the results. The dashed lines show the expected limit if the trident cross-section could be measured with 10\% or 30\% accuracy using 5~GeV neutrinos scattering on Argon.}
\label{fig:constraints2}
\end{figure}

Neutrino trident production has been searched for in several neutrino beam experiments. Both the CHARM-II collaboration~\cite{Geiregat:1990gz} (using a neutrino beam with mean energy of $E_\nu \sim 20$~GeV and a glass target) and the CCFR collaboration~\cite{Mishra:1991bv} (using a neutrino beam with mean energy of $E_\nu \sim 160$~GeV and an iron target) reported detection of trident events and quoted cross-sections in good agreement with the SM predictions,
\begin{align}
\sigma_{\rm{CHARM-II}}/\sigma_{\rm{SM}} = 1.58 \pm 0.57  \label{eq:CHARM}
 ~,\\
 \sigma_{\rm{CCFR}}/\sigma_{\rm{SM}} = 0.82 \pm 0.28  \label{eq:CCFR}~.
\end{align}
(Corresponding results from NuTeV can also be used albeit with some caution due to a rather large difference in the background treatment between the initial report \cite{Adams:1998yf} and the publication \cite{Adams:1999mn}.)
These results strongly constrain the gauged $L_\mu-L_\tau$ model, and more generally any new force that couples to both muons and muon-neutrinos. 
Implementing the phase space integrations that correspond to the signal selection criteria of CCFR and CHARM-II, we arrive to the sensitivity plots in
 Figs.~\ref{fig:constraints} and~\ref{fig:constraints2}.
Our results show that the parameter space favored by the muon $g-2$ discrepancy is entirely ruled-out above $\mzp \gtrsim 400\MeV$, proving the 
importance of neutrino trident production for tests of physics beyond the SM. 

{\em Other constraints and future possibilities.} 
As can be seen from Fig.~\ref{fig:constraints}, the region between $5 \lesssim \mzp \lesssim 50\GeV$ is 
independently constrained by searches for the SM $Z$ decay to four leptons at the LHC~\cite{CMS:2012bw,Aad:2014wra}. The bound obtained by recasting the ATLAS search~\cite{Aad:2014wra}, based on the full 7+8 TeV data set, extends down to $g^\prime\sim 10^{-2}$ at $\mzp\sim 10$ GeV. However, the sensitivity diminishes at low $\mzp$ because of the cuts employed in this specific LHC search, and in particular on the invariant mass of same flavor opposite sign leptons. The clear sensitivity of high-energy colliders to this region of parameter space motivates a dedicated search targeting the specific topology of an on-shell $\zprime$ emitted from the muonic decay of the $Z$ vector-boson and consequently decaying into a pair of muons. At quite low $\mzp$ a complication arises as the $\zprime$ becomes more boosted and the muons originating from its decay are more tightly collimated, forming a so-called ``lepton-jet''~\cite{ArkaniHamed:2008qp}.
Thus, low-mass leptonic $\zprime$ points to an interesting prospect of a search for events with two opposite-sign muons in addition to one muon-jet, altogether reconstructing the $Z$ boson. 

Searches at B-factories for four lepton events can also be sensitive to the low $\mzp$ region. A search by BaBar looked at the pair production of two narrow resonances, each decaying into a $\mupmum$ (or $e^+ e^-$) pair~\cite{Aubert:2009af}. While that search was optimized to an underlying two-body 
event topology, with two equal masses, rather than one resonance, we can use it to gain insight on the potential sensitivity of a  dedicated search of $\zprime$. Requiring the $\zprime$ to contribute less than 10 events in each, 100 MeV wide, bin of the $\mupmum$ invariant mass distribution shown in ref.~\cite{Aubert:2009af}, we  estimate a sensitivity to a coupling 
at the level of
$\gp \sim 2\times 10^{-2}$ for $\zprime$ masses in the range $0.5 \lesssim \mzp \lesssim 5\GeV$. Dedicated analyses of BaBar and Belle data, as well as future searches at Belle II might be able to probe couplings down to few$\times 10^{-3}$
over a wide kinematic window of $m_\zprime$, open for direct $\zprime$ production with subsequent decay to muon pairs.

\begin{figure}[t]
\centering
\includegraphics[width=0.41\textwidth]{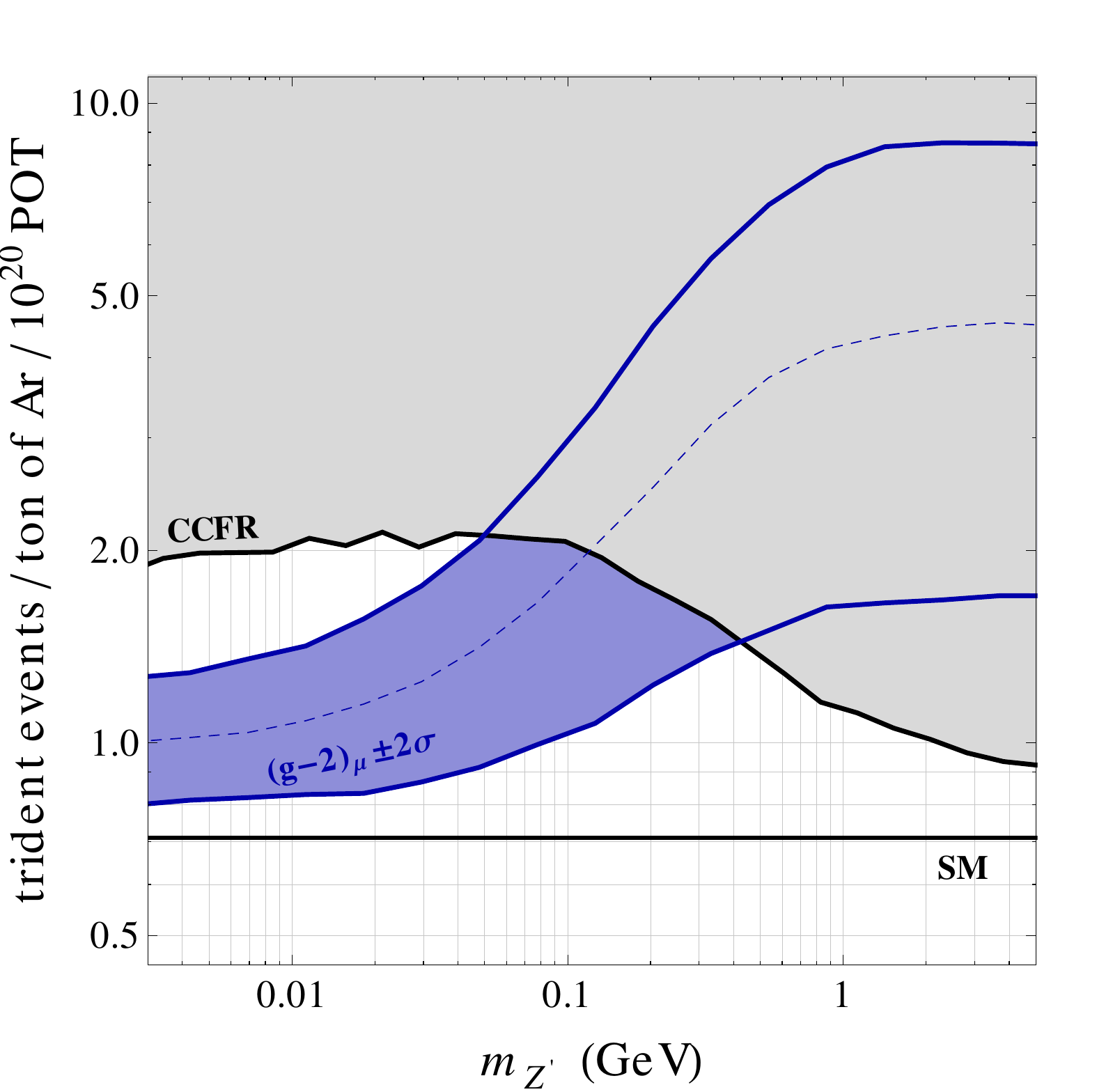}%
\caption{Expected number of trident events per ton of Argon and per $10^{20}$~POT at the LBNE near detector for a neutrino energy of $E_\nu = 5$~GeV as a function of the $\zprime$ mass. The horizontal line shows the SM prediction. The purple (dark grey) region corresponds to $\zprime$ masses and couplings that yield a contribution to the muon g-2 in the range $\Delta a_\mu = (2.9 \pm 1.8) \times 10^{-9}$. The light grey region is excluded by CCFR. 
}
\label{fig:LBNE}
\end{figure}

Perhaps even more interestingly, the low $\mzp$ region can be efficiently explored at the planned neutrino facility LBNE, with its lower energy and higher luminosity, as compared to past neutrino beam experiments. In Fig.~\ref{fig:LBNE} we show an estimate for the expected number of trident events per ton of Argon and per $10^{20}$ protons-on-target (POT) at the near detector at a LBNE-like run where for simplicity we set the neutrino energy to $E_\nu = 5\GeV$. For our estimate we use the expected charged current rates from~\cite{LBNE1} and the charged current cross sections from~\cite{Beringer:1481544}. With about one year of data (corresponding to $\sim 6 \times 10^{20}$~POT~\cite{Adams:2013qkq}) and a $\sim18$ ton Argon near detector setup~\cite{LBNE3}, we expect $\mathcal{O}(100)$ trident events in the region of parameter space favored by the muon g-2 anomaly with $\sim 30-100\%$ contribution from new physics. Needless to say, a more thorough study is needed before the precise sensitivity can be established. Nevertheless, these initial numbers suggest very favorable prospects for discovery sensitivity in this region of parameter space
of the leptonic force models.

{\em Acknowledgments.} We would like to thank T. Adams for useful correspondence and Eder Izaguirre and Brian Shuve for discussions. WA and SG would like to thank the SLAC theory group for hospitality and support. The research of 
WA was supported by the John Templeton Foundation. Research at Perimeter Institute is supported by the Government of Canada through Industry Canada and by the Province of Ontario through the Ministry of Economic Development \& Innovation. 

\bibliography{mu-tau_bib}

\end{document}